\documentclass{optica-article}
\journal{opticajournal} % for journals or Optica Open
\articletype{Research Article}
\usepackage{lineno}
\usepackage{xspace}
\usepackage{amsmath}
\usepackage{graphicx}
\usepackage{xcolor}
\usepackage{comment}
\usepackage{multirow}
\usepackage{rotating}
\usepackage{lineno}
\usepackage{graphicx}
 
\newcommand{\apriori}{\textit{a priori}\xspace}

\graphicspath{{./}{./figures/}}
\begin{document}

\title{Critique of optical coherence tomography in epistemological metrology}
\author{Yoshiaki Yasuno\authormark{*}}

\address{Computational Optics Group, University of Tsukuba, Tsukuba, Ibaraki 305-8573, Japan.}
\email{\authormark{*}yoshiaki.yasuno@cog-labs.org} 
\homepage{https://optics.bk.tsukuba.ac.jp/COG/}

\begin{abstract}
This treaties aims to introduce  a speculative framework to comprehend modern metrology.
In a classical sense, measurement measures physically existing quantities, such as length, mass, and time.
By contrast, modern measurement measures a wide spectrum of objects, for example, an imaging biomarker, which is a user-defined pattern.
Artificial-intelligence (AI)-based diagnosis is another example.
If we regard an optical coherence tomography (OCT) device equipped with diagnostic AI as a comprehensive ``measurement system,'' the object being measured is a ``disease,'' which is not a physical quantity but a concept.

To comprehend the wide range of modern measurements, we introduce a speculative, i.e.,  philosophical and theoretical, model called the ``epistemological-metrology model.''
In this model, we describe the act of measurement as cascading encoding-and-decoding processes.
In the encoding processes, three types of objects-to-be measured are considered, which include substance, existence, and concept.
Then we classify acts of measurement into ``sensing,'' ``understanding,'' and ``reasoning,'' which measure the substance, existence, and concept, respectively.

We note  that the measurements in the understanding and reasoning classes are constructive.
Namely, they proactively define the quantity-to-be-measured by the measurement modalities themselves.
A speculative method to warrant the relevance of such constructive measurements is presented.

We investigate several modern OCT-related measurements, including AI-based diagnosis, types of polarization sensitive OCT, and attenuation coefficient imaging, using our theoretical framework.
\end{abstract}
	
\section{Background: Modern conception of measurement}
The modern conception of measurement includes far broader fields than that in the classical sense.
In classical measurement, the quantities to be measured are quantities that are physical substances, such as mass, length, and time.
In the case of conventional, or classical, optical coherence tomography (OCT) \cite{Huang1991Science}, the quantity to be measured is back-scattering light intensity.
However, modern measurement systems measure a far broader spectrum of quantities.

An example is a clinical imaging device equipped with an automatic diagnostic function, such as an OCT device with glaucoma diagnostic software, for example Refs.\@ \cite{BurganskyEliash2005IOVS, Mwanza2016COO}.
The final measurement target of such a device is ``glaucoma,'' that is, a ``disease.''
A disease is not a physical entity but a concept.

A similar example is an ``imaging biomarker,'' which is a characteristic pattern in an image that correlates with a specific disease, for example Ref.\@ \cite{Sullivan2015Radiology}.
When a measurement system acquires an image and computationally obtains the imaging biomarker, the object being measured by such a system is a biomarker.
This is not a physical entity but a user-defined artificial quantity.

A more specific example can be found in polarization-sensitive (PS)-OCT \cite{deBoer2017BOE, Baumann2017AS}.
Some PS-OCT measures a quantity called the ``degree-of-polarization-uniformity'' (DOPU) \cite{Gotzinger2008OpEx}.
Although the DOPU is highly related to a physically defined quantity; degree of polarization or depolarization, DOPU and these quantities are not the same.
The DOPU is defined as a variance of Stokes vectors on a Poincar\'e sphere measured within a small spatial region in an image.
DOPU is not defined based on physical models but is a quantity arbitrarily defined by PS-OCT developers.
As metrology and measurement science progress, the spectrum of the object being measured will be further broadened and further exceeds the domain of physical entities.

\section{Complication in modern measurement}
Modern measurement systems are based on a variety of technologies, theories, and conceptions.
They cover, for example, physical phenomena (e.g., interference), physics-based signal processing (e.g., OCT reconstruction), image processing (e.g., tissue segmentation and imaging biomarker extraction), and image recognition (e.g., artificial-intelligence (AI)-based computer diagnosis). 
Although these technological concepts are used together, their mutual and theoretical relations are not well understood.
As we discuss in later sections, the lack of understanding has occasionally caused irrelevant uses of some of these concepts.
Additionally, this irrelevance will become more severe as modern measurement begins to cover wider conceptions. 

\section{Our approach to comprehend ``measurement''}
This treaties aims to present a speculative (philosophical) framework to comprehend modern measurement.
In this framework, conceptions of modern measurement are organized as a theoretical and speculative model called ``epistemological metrology.''
To induce this model, we start from a readily acceptable generalized model of measurement: the ``cascading encoding-decoding model.''
Then we identify the similarity between this model and the epistemology of Immanuel Kant \cite{Kant1787}.
By exploiting this similarity, we induce the ``epistemological-metrology model,'' which comprehends the conceptions of classical and modern measurements.

In epistemological metrology, we classify measurements into three classes: ``classical,'' ``semi-classical'' (or ``semi-cognitive''), and ``cognitive.''
Cognitive and semi-cognitive measurements are characterized by their constructive nature; that is, cognitive measurement is defined as a measurement methodology that defines, or constructs, the object-to-be-measured (OTBM) using the measurement methodology itself.
By contrast, classical measurement is measurement in a well-accepted sense, i.e., it measures quantities that were defined before the measurement methodology was established.

To further understand the conceptions of classical, semi-cognitive, and cognitive measurements, we introduce three types of OTBM including (1) substance, (2) existence, and (3) concept.
They are measured in the classical, semi-cognitive, and cognitive measurements, respectively.

We also show that some currently well-accepted measurements, including PS-OCT and AI-based diagnosis, are not classical but semi-cognitive or cognitive; that is, they construct (i.e., define) the quantities to be measured themselves.
We also note that cognitive measurement must be constrained by some means to maintain its relevance.
For this constraint, we introduce the idea of ``inevitability.''
We show that the idea of inevitability not only warrants the relevance of the measurement but also clarifies the difference between signals and artifacts in a speculative sense.

\section{Cascading encoding-decoding model}
\label{sec:edmodel}
\begin{figure}
	\centering\includegraphics[width=12cm]{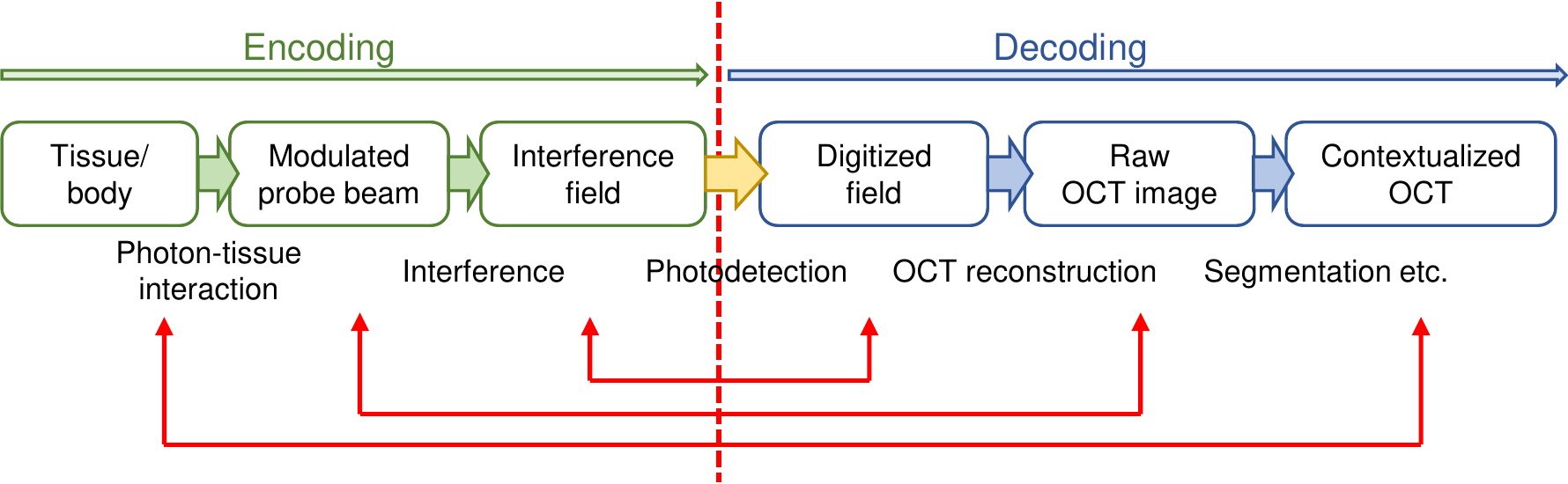}
	\caption{%
		Sequence of OCT measurement.
		The first half, from tissue to the interference field, can be  regarded as an encoding process, whereas the second half, from the digitized field to the contextualized OCT image, is regarded as a decoding process.
		The first and second halves show a symmetricity centered  at photo-detection.
		See the first two paragraphs of Section \ref{sec:edmodel} for details.}
	\label{fig:OctSequence}
\end{figure}
Measurement can be regarded as a sequential process of encoding and decoding. 
We denote this view of measurement as a ``cascading encoder-decoder model.''
Figure \ref{fig:OctSequence} exemplifies the measurement process of OCT.
The first object is biological tissue or the human body.
It interacts with the probe beam, which results in a modulated (and backscattered) probe beam.
This modulated probe beam interferes with a reference beam and forms an interference field.
This interference field is detected and digitized.
The digitized interference field is processed, and a raw OCT signal is yielded.
The raw OCT signal is further processed and a contextual image of tissue or the body is obtained.
The contextual image is exemplified by tissue-segmented OCT images, an attenuation coefficient (AC) image (e.g., Ref.\@ \cite{Vermeer2014BOE}), a DOPU image of PS-OCT \cite{Gotzinger2008OpEx}, optical coherence refraction tomography images \cite{Zhou2019NatPhoton, Zhou2022Optica}, etc.

We find that this sequential process is symmetric and centered at the inference-field-detection step (the dashed line in Fig.\@ \ref{fig:OctSequence}).
We also find that the left-hand side of the symmetric axis is a cascading encoding process from a physical entity (i.e., tissue), to a phenomenon, that is, the interference signal. 
The right-hand side is a corresponding decoding process that starts at the digitized interference field and reaches the image of the measurement target. 

\begin{figure}
	\centering\includegraphics[width=10cm]{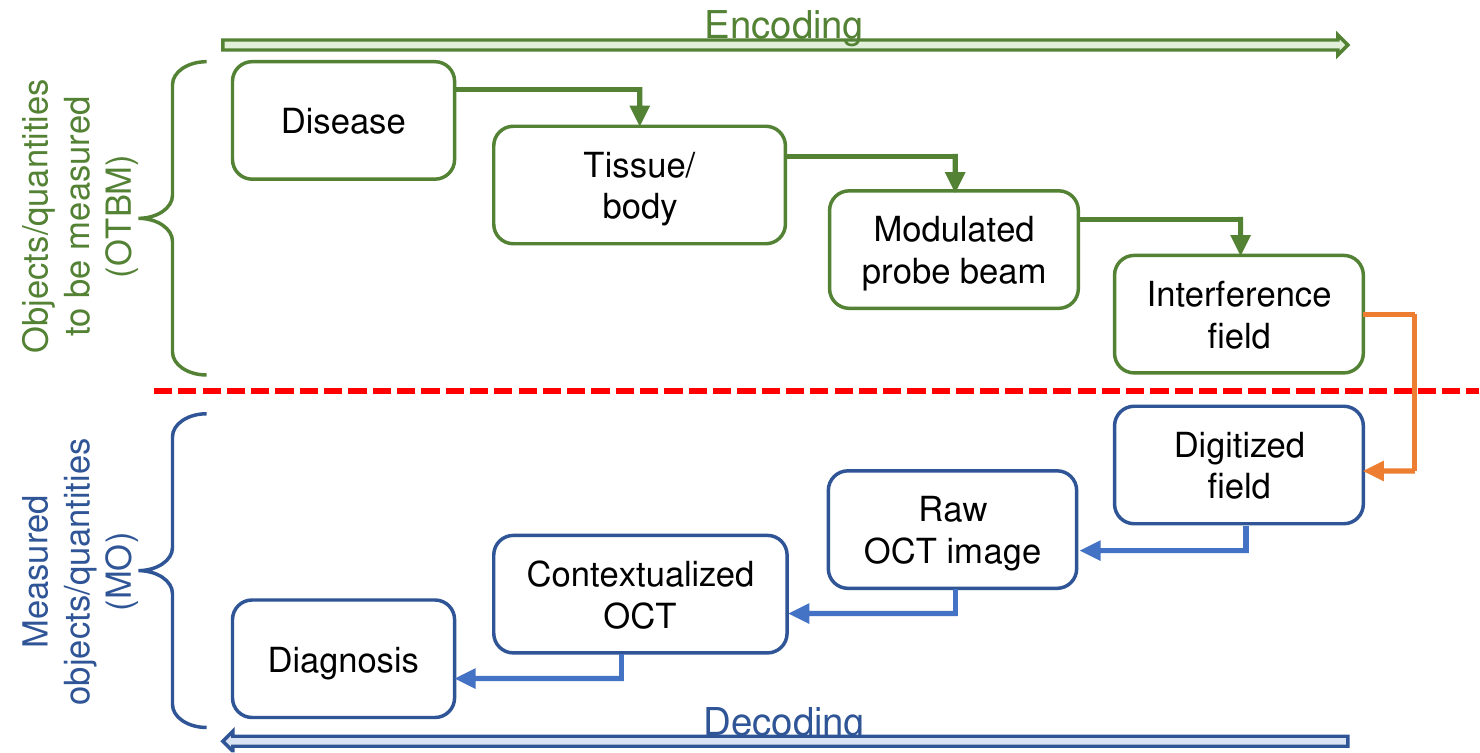}
	\caption{Measurement in its wide sense is depicted as a cascading encoding-and-decoding process.
		The objects in the upper half of the diagram are the objects-to-be-measured, whereas those in the lower half are the measured object/quantity.
		Objects at the same horizontal position correspond to each other.}
	\label{fig:cascading}
\end{figure}
This model of an imaging process becomes more comprehensible in the depiction in Fig.\@ \ref{fig:cascading}. 
The objects in the encoding process (upper half of the diagram) and corresponding objects in the decoding process (lower half of the diagram) are written at the same horizontal locations, which are referred to as ``levels.'' 
Additionally, we added more objects at the beginning and end of the process, that is, diseases and diagnosis.
A disease abnormalizes tissue and the diagnosis software of the measurement system provides a diagnosis using the abnormality of the contextualized image.

It is noteworthy that the distance between the objects at the same level indicates the similarity between the objects.
For example, the similarity between the interference signal and its digitized version is higher than that between the human body and its contextualized image. 

\section{Epistemology and metrology}
In this section, we note that the cascading-encoding-decoding model is similar to the epistemological model of Kant \cite{Kant1787}.
The Kant’s model  explains the act of human recognition as a sequence of three processes: sensing, understanding, and reasoning.

In the sensing process, the human cognitive system captures the phenomenon as it appears and passes it to the understanding process.
In this process, the sensed object is not yet contextualized.
This process is similar to the digitization of the interference field, in which the interference electric field is converted into a digital signal, but its meaning, that is, context, is not yet known.

The understanding process binds the sensed object to a particular meaning, such as the particular name of the phenomenon.
This process is similar to OCT image reconstruction and tissue segmentation.
The meaningless interference signal is properly processed, and a meaningful, on the other word, ``contextualized,’’ OCT image is generated.

The final process, that is, reasoning, leads us to the ``concept behind the contextualized phenomenon.''
This concept is purely conceptional and its existence or nonexistence cannot be proven.
The ``god'' is an example of such a concept in Kant's philosophy.
The reasoning process is similar to ``diagnosis'' in our measurement.
In the diagnosis process, we measure the disease, that is, the reason for tissue abnormality.
The disease is assumed to be the cause of abnormality and is purely conceptional.
It is also noteworthy that, in reality, the disease did not abnormalize the tissue, but we defined a concept of the disease to explain the cause of abnormality.

Considering the similarity between Kant's model and our measurement model, we find that the OTBM are more physical on the left-hand side of the diagram (Fig.\@ \ref{fig:cascading}) and become more conceptual, or metaphysical, on the right-hand side.
It is also noteworthy that, on the right-hand side, the object in the upper half of the diagram is more fundamental, whereas on the left-hand side, the object in the lower half is more fundamental; that is, the interference electric field  is more fundamental than the digitized signal, that is, the electric field `` rules'' the digitized signal.
By contrast, the diagnosis, that is, the recognized disease, is more fundamental and important than the conceptual ``disease.''
We provide details regarding this point in the next section.

\section{Classical and cognitive measurements}
\label{sec:classicalAndCognitive}
\begin{figure}
	\centering\includegraphics[width=12cm]{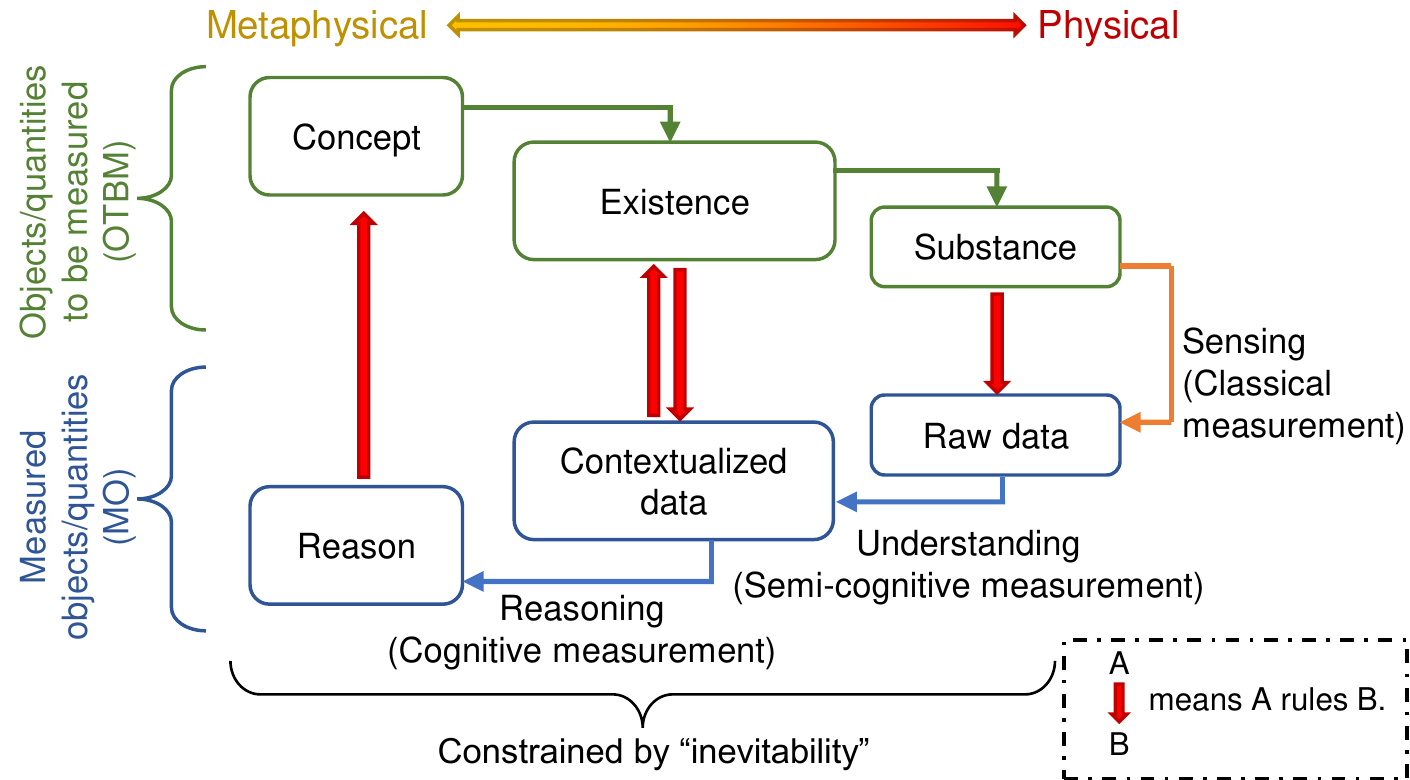}
	\caption{Model of ``epistemological metrology.''
		The objects-to-be-measured (OTBM) are classified into substance, existence, and concept according to the ruling relationship between the OTBM and the measured object (MO).
		Acts of measurement are classified into sensing, understanding, and reasoning, and they correspond to classical, semi-cognitive, and cognitive measurements, respectively.}
	\label{fig:emModel}
\end{figure}
\begin{table}
	\caption{The key conceptions appeared in the epistemological metrology model.
		The object-to-be-measured (OTBM) is classified into the substance, existence, and concept (first row), based on the idea of ruling (second row).
		The measurement acts are then classified as sensing, understanding, and reasoning, respectively for the substance, existence, and concept.
		These measurement acts are recognized as classical, semi-cognitive (semi-constructive), and cognitive (constructive), respectively.}
	\label{tab:concepts}
	\centering\includegraphics[width=12cm]{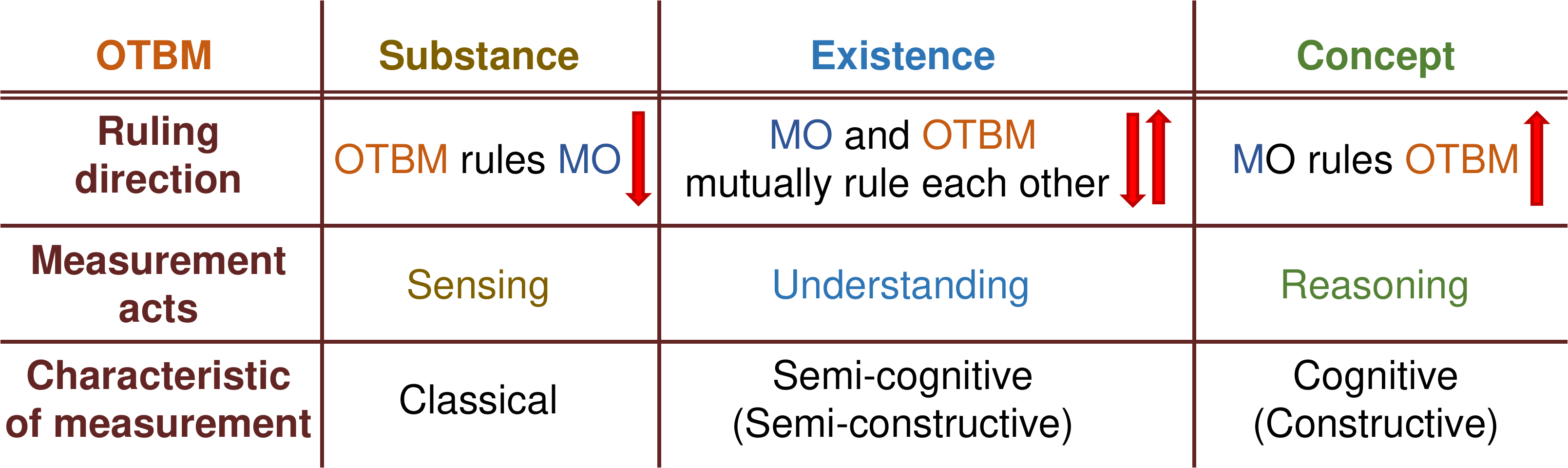}
\end{table}
By exploiting the similarity between the epistemology and the measurement, we can define a ``epistemological metrology model,'' as depicted in Fig.\@ \ref{fig:emModel}.
Table \ref{tab:concepts} also summarizes the key conceptions appeared in this model.
Here the first point is that the OTBM and the measured object/quantity (MO) are explicitly distinguished.
In the diagram, the OTBM are appeared in the upper side, while the MOs are appeared in the lower side.
The second point is that the OTBM are classified into three types: (1) substance, (2) existence, and (3) concept. 

Substance is a physical quantity or entity, such as an electric field.

Existence is an object that is definitively analogized from the measurement result of existence.
Contextualized OCT images, such as AC, DOPU, and tissue-segmented images, are the measurement results of existence.
A raw OCT image is the measurement result of substance in one perspective, or existence in another perspective.
Namely, if we consider that the OCT image represents the electric field distribution in a sample, OCT measures the substance.
On the other hand, if we consider the OCT image represents ``the scattering ability'' of the tissue, not the scattered electric field itself, then OCT measures the existence.

The concept is an object that is arbitrarily defined to explain the measured existence.
Disease is an example of a concept.

Based on the classification of the OTBM, we can define three levels of measurements.
The most fundamental level of measurement is sensing, which measures physical existence (rightmost of the diagram in Fig.\@ \ref{fig:emModel}).
The second level of measurement is understanding, which measures existence.
Existence is more contextualized than substance (middle of the diagram).
The final level of measurement is reasoning, which measures a concept to explain the reason behind the measured existence (leftmost).
For example, an interference field (substance) is sensed by a photo-detector and is understood as a contextualized OCT image (existence) and is finally reasoned as a disease (concept).
The further to the left the quantity is in the diagram, the more contextualized.

To further understand the levels of measurement, we introduce the idea of ``classical'' and ``cognitive'' measurements.
A cognitive measurement is characterized by its constructive nature; that is, cognitive measurement proactively defines the OTBM.
So, we also denote the cognitive measurement as ``constructive measurement.''
Reasoning is a cognitive (constructive) measurement.
In our example, the disease is ``defined,'' i.e., the concept of the disease is constructed, by being diagnosed; that is, the observation (diagnosis) ``rules '' the object (disease), as shown by the red arrow in the diagram.

By contrast, classical measurement measures a predefined quantity, such as an electric field.
In this case, the predefined physical quantity (i.e., OTBM) rules the measured quantity (MO).
Sensing is a classical measurement.

The understanding process is classified as ``semi-classical'' or equivalently ``semi-cognitive (semi-constructive).''
The measured quantity is highly related to the predefined physical object (or phenomenon), but the measured quantity is contextualized by the measurement.
Hence, the OTBM and the MO mutually rule  each other.

\section{``Inevitability'' constrains cognitive measurements} 
An immediate question that arises about cognitive and semi-cognitive measurements is ``if the OTBM is defined by the act of measurement itself, how can the measured quantity be meaningful?''
In cognitive measurement, anything could, in principle, be arbitrarily defined as an OTBM.
Hence, it should be constrained in some manner.
To constrain cognitive and semi-cognitive measurements and make them meaningful, we introduce the idea of the ``inevitability of the measurement.''
Inevitability explains why the quantity ``should be measured.''
We define two types of inevitabilities: pragmatic and transcendental, as summarized in Table \ref{tab:inevitability}. 

\begin{table}
	\caption{
		Two types of inevitabilities warrant the cognitive and semi-cognitive (constructive and semi-constructive) measurements.
	}
	\label{tab:inevitability}
	\centering\includegraphics[width=12cm]{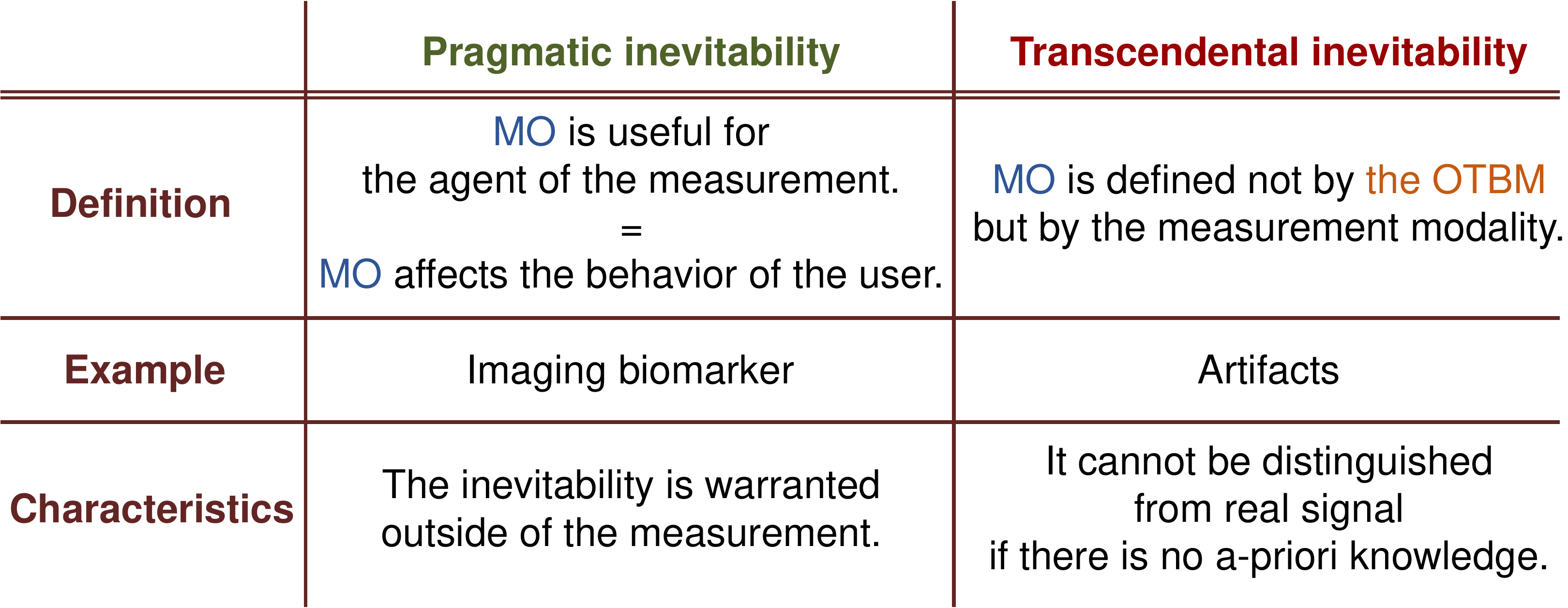}
\end{table}

The terminology of ``pragmatic inevitability'' was inspired by the philosophical school of ``pragmatism'' \cite{James1907Pragmatism}.
In this philosophy, the proposition is thought to be meaningful only if it is effectual.
Based on a similar idea, we regard the OTBM to be ``pragmatically inevitable'' if the measured object is meaningful in any context; that is, pragmatic inevitability is defined based on the utility of the measurement.
For example, a ``disease'' is pragmatically inevitable if it can be used to make a clinical decision or develop an effective therapy.
Similarly, a segmentation result is pragmatically inevitable if it can be used for the next step of the measurement, that is, diagnosis.

The second type of inevitability is ``transcendental inevitability,'' which is inspired by Kant’s transcendental, or metaphysical, philosophy.
In this philosophy, some transcendental objects, such as the soul or a god, are thought to be inevitably recognized.
However, this recognition is not based on the existence of the object but purely based on the recognition system of the human being.
By adopting this concept, we define a transcendentally inevitable OTBM as follows: what appears in the measurement results but is originated from the measurement system itself instead of the measured sample.
Namely, the transcendentally inevitable object is the ``artifact,'' which inevitably appears in the measurement results because of the property of the measurement system. 

One important characteristic of the transcendentally inevitable object, that is, the artifact, is that it can be distinguished from the signal only if we have \apriori knowledge about the sample and/or measurement system.
For example, ``fixed pattern noise'' and/or the ``mirror image'' of OCT can be recognized as an artifact only if we have known that the sample does not have such a structure and the principle of OCT can create such pseudo-signals.

\section{Critique of OCT in epistemological metrology}
By exploiting the epistemological metrology model, we can preclude  some irrelevant directions of OCT research.
We provide examples from AI-based diagnosis, PS-OCT, and AC measurement.

\subsection{AI-based diagnosis}
AI-based diagnosis is ``reasoning,'' and hence, cognitive (i.e., constructive).
Some diagnostic AIs were designed to mimic the established classification systems of clinical findings, such as the classification of the optic nerve head.
If the training target of AI is human-classified labels, then the attempt of AI-based classification could, in principle, be cyclic.
This is because both the training target, that is, the classification results of human experts, and the AI are in the lower half of the diagram in Fig.\@ \ref{fig:emModel}.
In practice, this type of AI can be meaningful as long as  the accuracy of the AI’s inference is lower than the inter-human expert agreement.
However, if we attempt to increase accuracy more than the agreement, the AI does not lead us closer to the real disease, which is in the upper half of the diagram.
Namely, the pragmatic inevitability of such over-learning is low.

We can consider actively defining the disease based on the AI-based diagnosis.
%, for example, as shown in Ref.\@ \cite{Panda2022AJO}.
For example, we can train the AI to provide the best responding treatment using clinical observations as input.
In this case, the concept of the disease is defined by the AI because it can maximize clinical intervention.

\subsection{Degree-of-polarization-uniformity (DOPU)}
Another example can be found in DOPU-related research.
Low DOPU values are believed to be correlated with melanin \cite{Baumann2012BOE}; hence, a low DOPU region is collocated with retinal pigment epithelium (RPE) \cite{Gotzinger2008OpEx}.
However, DOPU measurement is semi-cognitive.

Historically, DOPU was ``defined'' to highlight RPE \cite{Gotzinger2008OpEx} after it was found that RPE exhibits random phase retardation patterns \cite{Pircher2004OpEx}.
Hence, there is no logical inevitability that relates DOPU to melanin.

This logical speculation implies that ``proving low DOPU is inevitably associated with melanin'' is not really a reasonable research direction, but defining modified versions of DOPU or similar polarization randomness metrics \cite{Makita2014OL, Lippok2015OL, Yamanari2016BOE} that are sensitive and specific to the tissue- or disease-of-interest can be more reasonable. 
This point also implies a fundamental similarity between DOPU and the idea of an imaging biomarker.

\subsection{Attenuation coefficient (AC) imaging}
AC imaging is another example of semi-cognitive measurement.
AC is the attenuation rate of the OCT signal or the back-scattering of the probe beam \cite{Vermeer2014BOE, Vermeer2012IOVS}.
Several attempts have been made to correct the confocal effect to achieve a more ``accurate'' AC measurement \cite{Smith2015ITMI, Kuebler2021BOE}.
We may find that if AC is defined as described above, the confocal effect is a part of the attenuation source, and hence should not be corrected.

This contradiction can be explained by considering the semi-cognitive nature of AC measurement; that is, AC measurement, in reality, is not intended to measure the signal attenuation rate but to measure ``some quantity being sensitive to tissue abnormalities;'' that is, AC measurement constructs the quantity-to-be-measured to meet the purpose of the measurement.
Hence, despite AC measurement seemingly being a measurement of physical quantity, it is semi-cognitive and its inevitability is pragmatically warranted.

AC is meaningful as far as it can be used to detect tissue abnormalities; hence, the definition of AC can be modified, for example, corrected for the confocal effect, to be more suitable for this purpose.

\section{Summary}
\begin{figure}
	\centering\includegraphics[width=8cm]{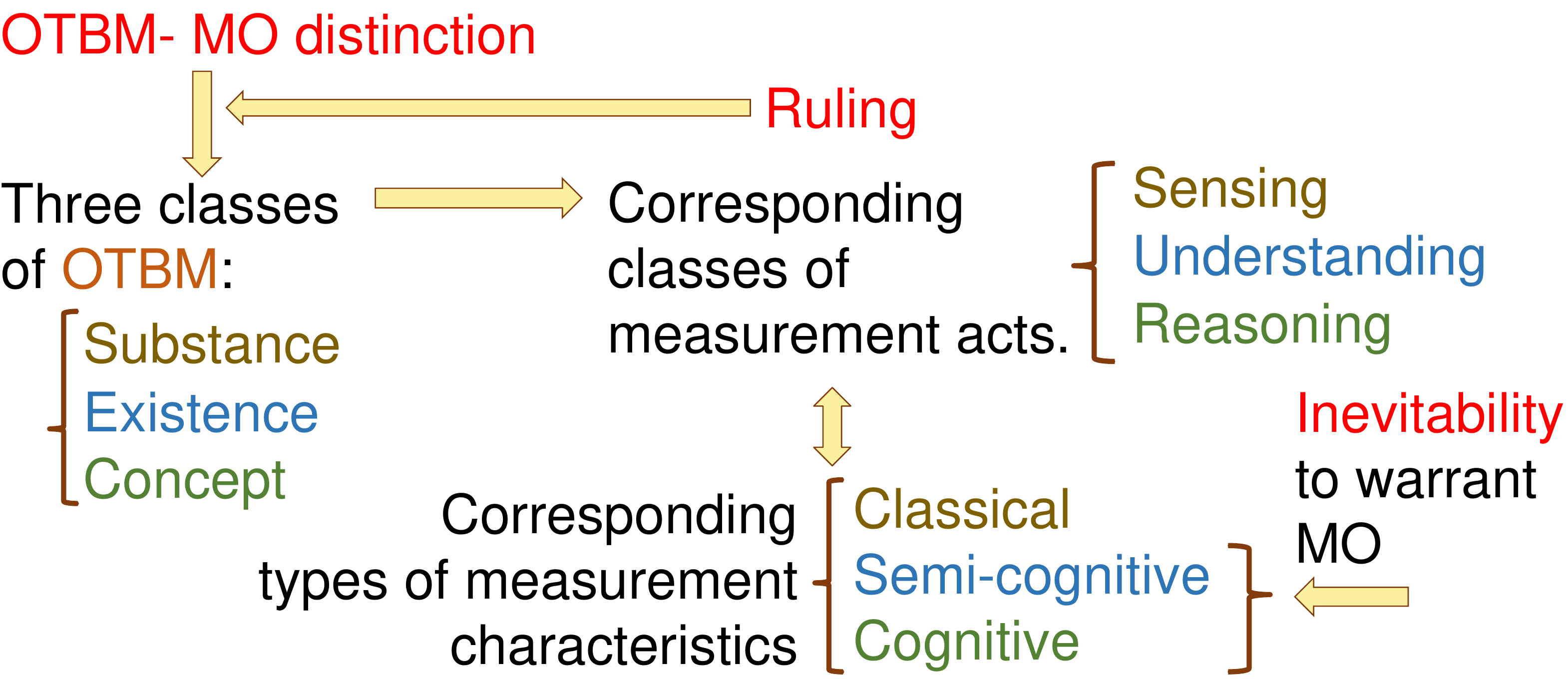}
	\caption{
		The summary of the logic structure on this treaties.
		Three primary concepts are highlighted by red.
		And other concepts are defined from these primary concepts.
	}
	\label{fig:logicDiagram}
\end{figure}
The logic structure of this treaties are summarized in a logic diagram shown in Fig.\@ \ref{fig:logicDiagram}.
In this treaties, we introduced a speculative model of wide-sense measurement called epistemological metrology to comprehend classical and modern measurements.
This model based on three key concepts: (1) the distinction between object-to-be-measured (OTBM) and measured-object (MO), (2) the ruling direction between the OTBM and the MO, and (3) inevitability of measurement.

Based on the ruling direction, we classified the OTBM into three types: substance, existence, and concept.

We classified the measurement measuring these objects as classical, semi-cognitive, and cognitive, respectively.
The latter two are characterized by their constructive nature, that is, they define the measured quantity as a part of the measurement.
Although the semi-cognitive and cognitive (i.e., constructive) measurements are not well constrained in a classical sense, the relevance of them can be warranted by the inevitability.
We found that several well-accepted OCT-based measurements are cognitive or semi-cognitive.
Additionally, we showed that this model is useful for evaluating the relevance of research directions in measurement science.

Although this treaties discussed only OCT and its related measurement modalities, the epistemological metrology model could be used to comprehend other measurement modalities.

\section*{Funding}
Core Research for Evolutional Science and Technology (JPMJCR2105);
Japan Society for the Promotion of Science (21H01836);

\section*{Disclosures}
Yasuno: Yokogawa Electric Corp. (F), Sky Technology (F), Nikon (F), Kao Corp. (F), Topcon (F), Tomey Corp (F).

\bibliography{reference.bib}

\end{document}